\documentclass[twocolumn,prd,floatfix,nofootinbib]{revtex4}

\usepackage{amsmath}
\usepackage{graphicx}
\usepackage{dcolumn}
\usepackage{bm}
\usepackage{amssymb}
\usepackage{latexsym}

\def\be{\begin{equation}}
\def\ee{\end{equation}}
\def\ba{\begin{eqnarray}}
\def\ea{\end{eqnarray}}

\begin{document}

\title{Scalar Perturbations Through Cycles}

\author{Zhi-Guo Liu\footnote{Email: liuzhiguo08@mails.gucas.ac.cn}}
\author{Yun-Song Piao\footnote{Email: yspiao@gucas.ac.cn}}

\affiliation{College of Physical Sciences, Graduate School of
Chinese Academy of Sciences, Beijing 100049, China}

\begin{abstract}

We analytically and numerically investigate the evolutions of the
scalar perturbations through the cycles with nonsingular bounce. It
is found that the amplitude of the curvature perturbation on large
scale will be amplified cycle by cycle, and the isocurvature
perturbations also obtain an amplification, but the rate of its
amplification is slower than that of curvature perturbation, unless
its coupling to the metric perturbation is not negligible.

\end{abstract}

\maketitle

\section{Introduction}

In recent decades, inflationary cosmology has achieved a lot of
successes, namely, it solved the homogeneity, isotropy, flatness
problems and offered a noble mechanism of generating a nearly
scale invariant spectrum of primordial fluctuations. However, as
the inflationary cosmology has singularity problem, it still
deserves a further investigation on relevant issues.

A cosmological cyclic scenario was proposed long time ago
\cite{Tolman}. In these years it has become popular \cite{STS},
which offered a new insight into the origin of observable
universe. The inflationary model bases on the idea that space and
time had a beginning when the universe had nearly infinite
temperature and density. In the cyclic model, the universe evolves
cyclically in time, undergoing endless expansions and
contractions, cooling and heating, in which the density and
temperature remain finite through these cycles.

There have been many studies on cyclic models
\cite{BD},\cite{KSS},\cite{Piao04},\cite{Lidsey04},\cite{Erick},\cite{CB},\cite{Xiong},\cite{Xin},\cite{Biswas},\cite{Cai0906},\cite{NB},\cite{Cai1108},
in which it is generally assumed that the cyclic universe is
homogeneous cycle by cycle. However, when the perturbations are
considered, the case might be
altered\cite{Piao0901},\cite{Piao1001},\cite{Zhang2010bb}. In
general, the amplitude of curvature perturbation on large scale will
increase in every contracting phase on large scale, however it will
be nearly constant in the expanding phase. Thus the net result of
one cycle is that the amplitude of curvature perturbation is
amplified. This amplification will possibly continue cycle by cycle.
In certain sense, the amplification of curvature perturbation might
eventually destroy the homogeneity of background. The effect of the
increasing of other perturbations on cyclic universe has been also
studied, e.g. anisotropic pressures \cite{BY}.

Since the perturbation modes is continuously increasing through
cycles, the reliability of some models of cyclic universe might be
required to reevaluate. In the model building of a consistent cyclic
universe, the perturbation should not only satisfy the requirement
of the observable universe, but also is not expected to destroy the
homogeneity of background. Thus the study of the evolutions of
cosmological perturbations in the model of cyclic universe is
significant.

In this paper, we firstly recheck the evolution of curvature
perturbation on large scale both analytically and numerically. It is
proved that the curvature perturbation and the metric perturbation
will be amplified cycle by cycle. Then we calculate the evolution of
the perturbation of a light scalar field through cycles, which
contributes the isocurvature perturbation. It is found that the
isocurvature perturbations also obtain an amplification, but the
rate of its amplification is slower than that of curvature
perturbation, unless its coupling to the metric perturbation is not
negligible.

\begin{figure}\label{fig:a}
\includegraphics[width=7.0cm]{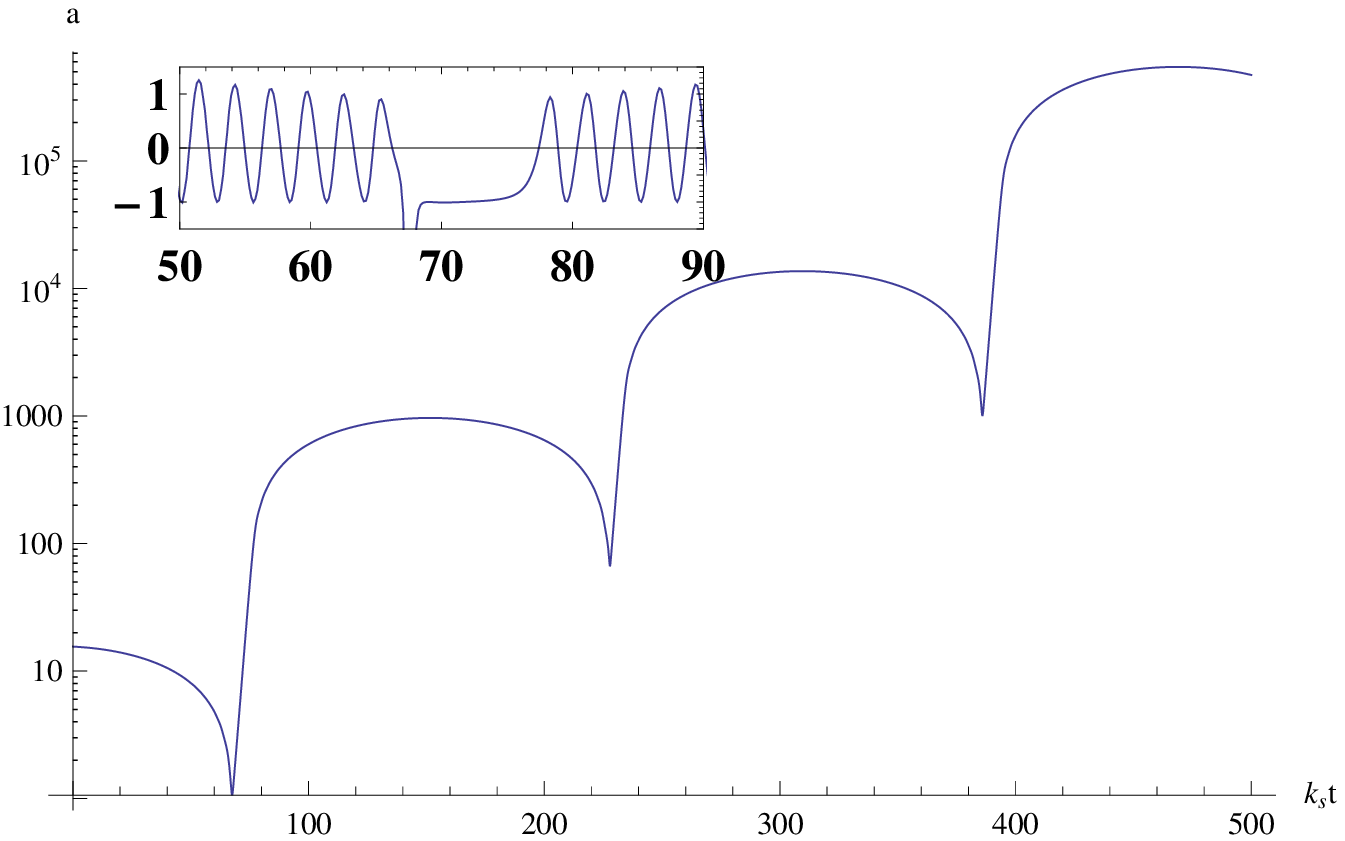}
\includegraphics[width=7.0cm]{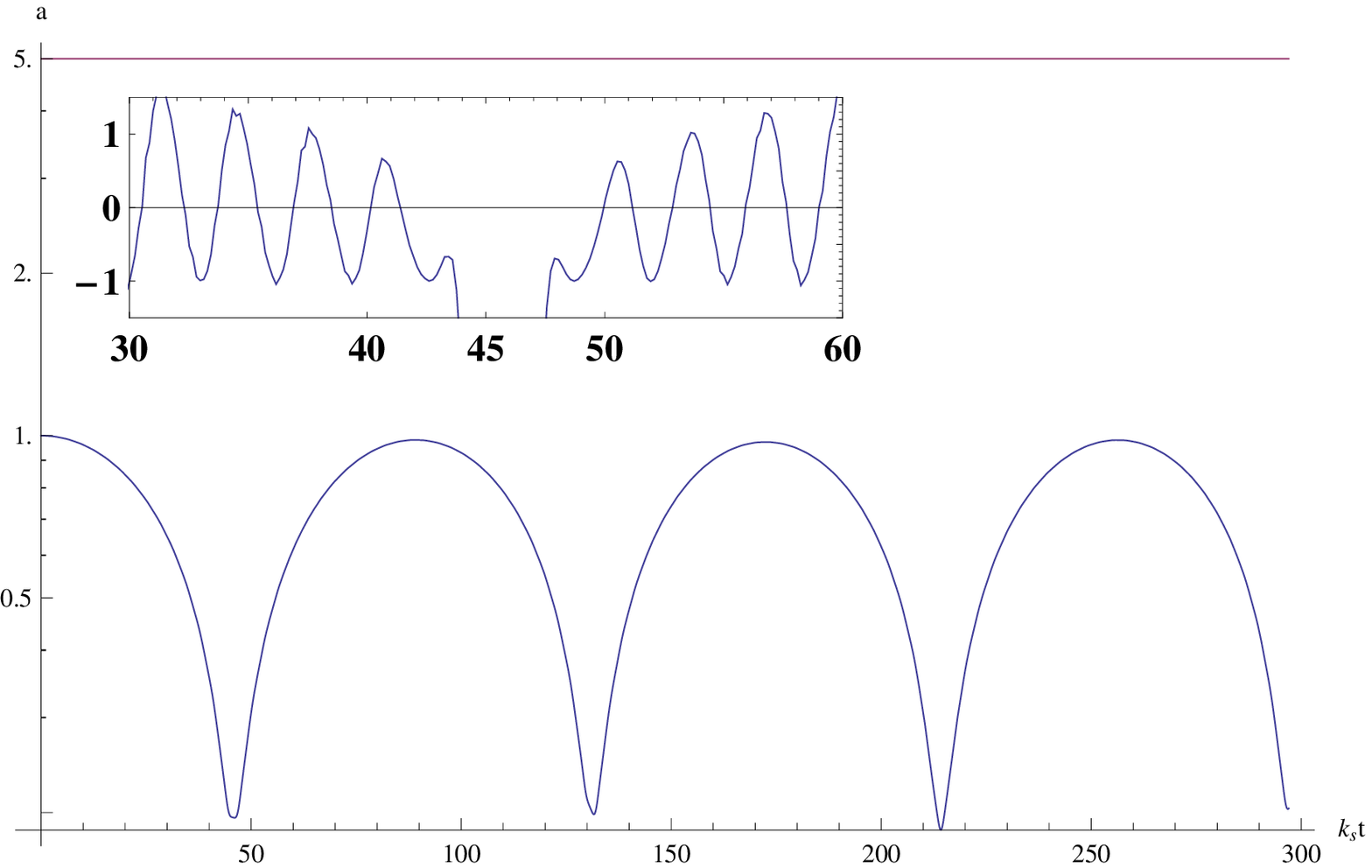}
\caption{\label{fig:a} The upper panel is the evolution of scale
factor in the model with increasing cycles. The lower panel is the
evolution of scale factor in the model with equal cycles. The time
is scaled by $k_s$ which is an adjustable parameter to make the
models be able to be consistent with the time scale of the
observable universe. The insets are the evolution of $\omega$
correspond to every cyclic model around the bounce time.}
\end{figure}

The outline of paper is as follows. In section II, we introduce
the models of cyclic universe, which will be used in following
sections, and the basic equations of perturbations. In section
III, we show the evolutions of the curvature perturbation on large
scale through cycles, and also the metric perturbation, which will
be used in section IV, in which we study the behavior of the
isocurvature perturbation on large scale without or with the
effect of the metric perturbation. The final section is the
discussion.


\section{The setup and The basic equation }

\begin{figure}[htbp]
\includegraphics[scale=0.6,width=6.0cm]{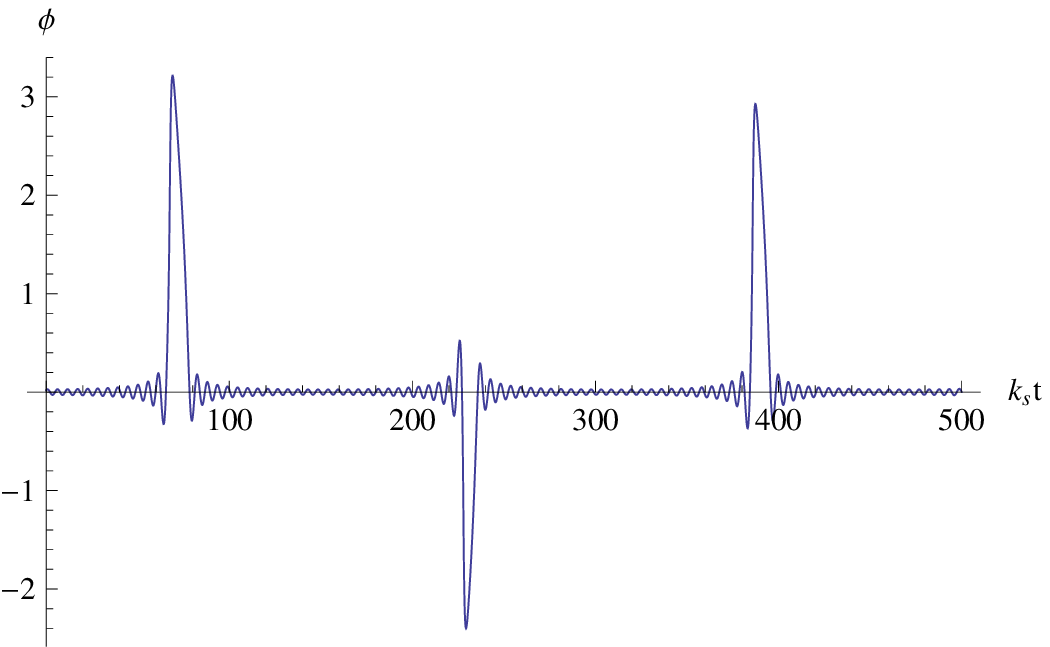}
\includegraphics[scale=0.6,width=6.0cm]{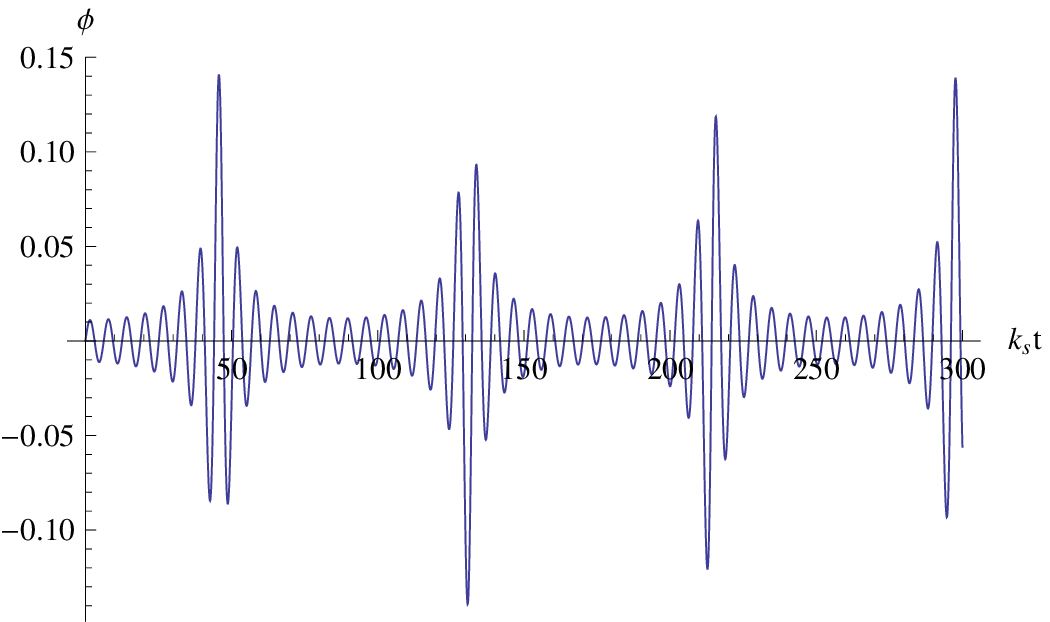}
\caption{The evolution of background field $\phi$. The upper panel
and the lower panel are these of the models in Fig.\ref{fig:a}
corresponding with increasing cycles and equal cycles, respectively.
} \label{fig:phi}
\end{figure}

In this section, we will introduce the models of cyclic universe,
which will be used in following sections, and the basic equations
of perturbations. We will regard the turnaround time as the
beginning of a cycle, which is $t_T^j$ in the $j^{th}$ cycle. In
each cycle the universe will sequentially experience the
contraction, bounce, and expansion, and then arrive at the
turnaround, which signals the end of a cycle. The beginning and
the end of the contracting phase and the expanding phase in the
$j^{th}$ cycle are $t_{Ci}^j$,
$t_{Ce}^j$, $t_{Ei}^j$ and $t_{Ee}^j$, respectively, and the bounce is 
$t_B^j$. Throughout this paper, ``the model with equal cycles"
means that the scale factor is periodically equal, while ``the
model with increasing cycles" means that the maximal value of the
scale factor or the scale of cycle is cyclically increasing.

The upper panel in Fig.\ref{fig:a} is the evolution of scale factor
of the model with increasing cycles, e.g. see \cite{Xiong} for
details. The models with increasing cycles have been obtained in
e.g.\cite{BD},\cite{KSS},\cite{Biswas}, in which the entropy is
increased, however, in our paper it results from a rapid expanding
period after the bounce in each cycle, the corresponding observable
signals have been studied in e.g.
\cite{Piao0308},\cite{Piao0310},\cite{Piao0501}. The lower panel is
that with equal cycles. We also give the evolution of background
field $\phi$ in Fig.\ref{fig:phi}.

In Fig.1, the nonsingular bounce is implemented by applying a
ghost field. In general, the introduction of ghost field is only
the approximative simulation of a fundamental theory below certain
physical cutoff. In this sense, its appearance is only an artefact
of this approximation. There have been lots of studies on how the
perturbations pass through such a nonsingular bounce
\cite{AW},\cite{BV},\cite{Cai0704},\cite{Cai0810},\cite{CK}.
Recently, the nonsingular bounce has been obtained in higher
derivative theories of gravity \cite{BMS},\cite{BKM}, or Galileon
field \cite{Qiu1},\cite{Vikman1}, which are ghost free, and others
\cite{Cai1104},\cite{Qiu10}, and also earlier
e.g.\cite{CCM},\cite{TBF}.

We will investigate the evolutions of the scalar perturbations
through cycles, and will not involve the details of model
building. The models in Fig.\ref{fig:a} will serve the purpose of
numerical simulations in the calculations of perturbations. We
shall work in longitudinal gauge in which if there is not the
anisotropy only the metric perturbation $\Phi$ exist. The
perturbation equations are \cite{Sasaki96},\cite{L99},\cite{GWBM}
 \ba \delta\ddot\phi_i & + &
3H\delta\dot\phi_i-\frac{\nabla^2}{a^2}\delta\phi_i+\sum_j
V_{,ij}\delta\phi_j\nonumber\\ & = & 4\dot\Phi\dot\phi_i-2\Phi V_{,i},\label{u1}\\
-3H\dot\Phi &+& (\frac{\nabla^2}{a^2}-3H^2)\Phi \nonumber\\ & =&
4\pi
G\sum_i[\dot\phi_i\delta\dot\phi_i-\dot\phi_i^2\Phi+V_i\delta\phi_i],\label{u2}
\\ \dot\Phi+H\Phi & = & 4\pi G\sum_i\dot\phi_i\delta\phi_i,\label{u3}\ea
where $V_{,i}$ denotes the derivative of the scalar field
potential with respect to $\phi$. In principle, the modifications
implementing bounce are only reflected on the perturbation modes
at UV scale. While the perturbation modes we observed in today's
cosmological observations are mainly in far infrared regime, thus
this set of equations are trustable.

We, with this set of equations and the background evolutions in
Fig.1, will analytically discuss and numerically simulate the
evolutions of the curvature perturbation and the isocurvature
perturbation on large scale in the following sections.

\section{The evolution of curvature perturbation }

We will recheck the evolution of the curvature perturbation. The
background field is $\phi$, and other fields contributing the
isocurvature perturbation are neglected. Thus the curvature
perturbation is $\zeta= \Phi+{H\over {\dot \phi}}\delta\phi$. The
equation of curvature perturbation in the momentum space is
\cite{Muk},\cite{KS},\cite{Mukhanov}
 \be
u_k^{\prime\prime} +\left(k^2-{z^{\prime\prime}\over z}\right) u_k
= 0 ,\label{uk} \ee where $u_k=z\zeta_k$ and $z={a{\dot \phi}/H}$,
the prime denotes the derivative for the conformal time $\eta$.

When the perturbations are deep inside the Hubble scale $k^2 \gg
z^{\prime\prime}/z$, the solution of perturbation is $ u_k\sim
\frac{1}{\sqrt{2k}}e^{-ik\eta}$. While the perturbations are
outside of Hubble scale $k^2\ll z^{\prime\prime}/z$, Eq.(\ref{uk})
has general solution $u_k$, which gives e.g.\cite{Mukhanov}, \be
\zeta_k \sim C_{\zeta,2}\int{d\eta\over
z^2}+C_{\zeta,1},\label{c12}\ee where $C_{\zeta,1}$ and
$C_{\zeta,2}$ are constant only dependent of $k$. The perturbation
spectrum will be determined by which one of both terms is
dominated.

In general, for the contraction phase \be a\sim (t_B-t)^{n},
\label{ac}\ee where $t<t_B$ and $n$ is constant. Thus during the
contraction with $n>{1\over 3}$, the amplitude of $\zeta$ will be
dominated by the $C_{\zeta,2}$ mode,
\begin{eqnarray} \zeta_k \sim \int{d\eta\over z^2}\sim \left(t_B-t\right)^{1-3n}, \label{zeta}\end{eqnarray}
which is increasing. While for the expanding phase $a\sim t^n$,
$\zeta_k$ is constant. This implies that for a cycle of cyclic
universe during the contraction $\zeta_k$ is increased on large
scale, up to the end of contracting phase in corresponding cycle,
while during the expansion it becomes constant. Thus the net result
is that $\zeta_k$ on large scale is amplified, which is inevitable.
Here, $n$ is constant only for the convenience of discussion.
However, around the bounce, the case can not be described simply as
Eq.(\ref{ac}), the qualitative result that the curvature
perturbation on large scale is amplified is universal.

\begin{figure}[htbp]
\includegraphics[scale=0.6,width=7.0cm]{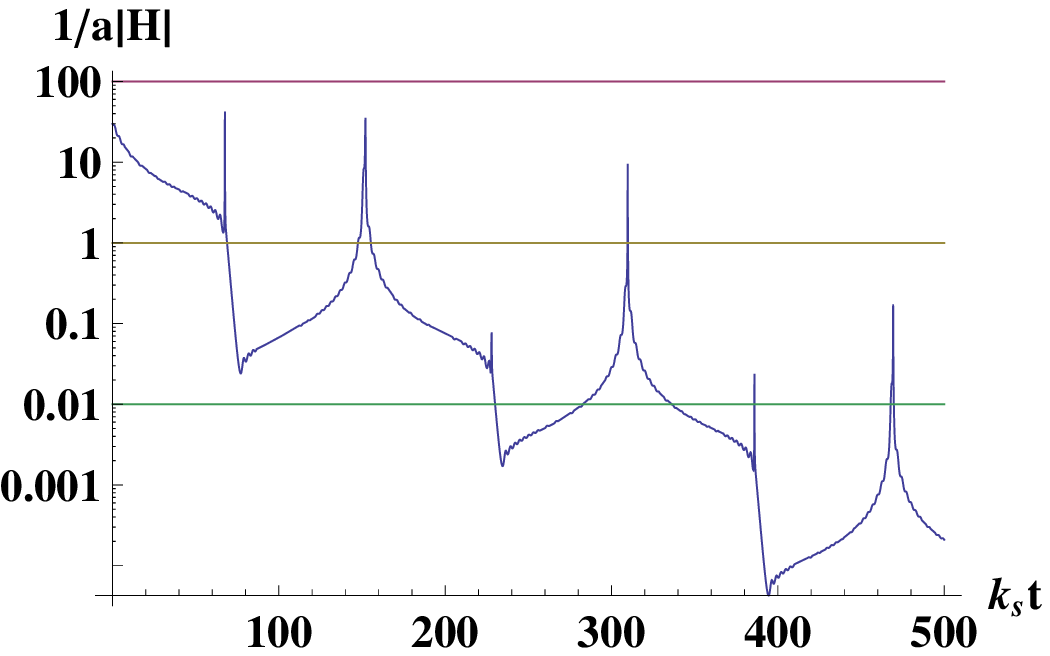}
\includegraphics[scale=0.6,width=7.0cm]{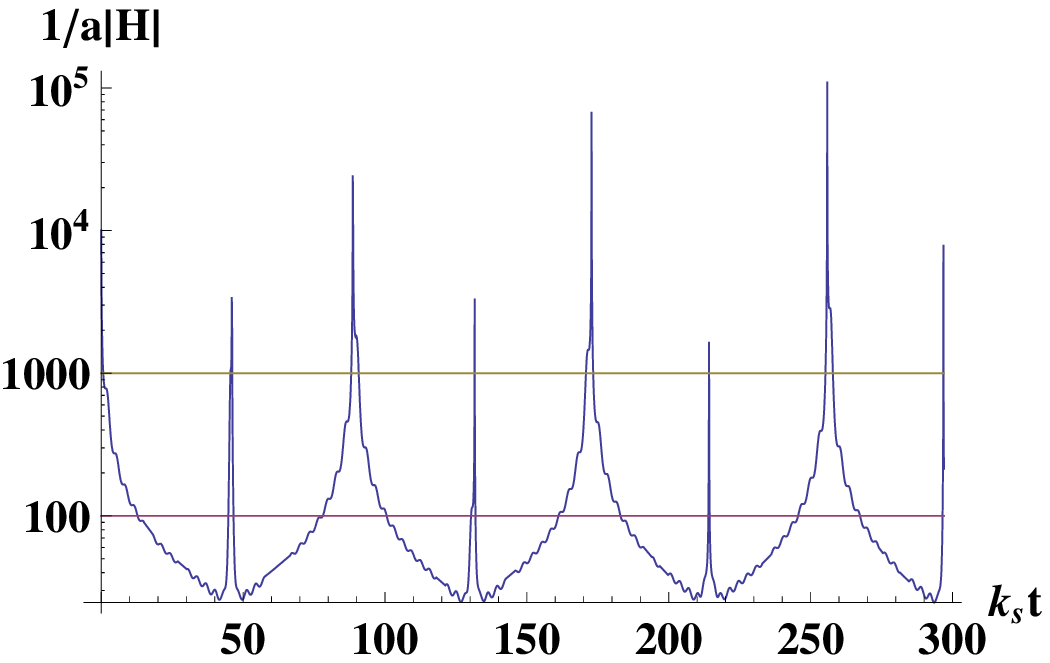}
\caption{ The evolutions of the comoving Hubble radius ${1\over
a|H|}$ and the perturbation modes are plotted for the models with
increasing cycles and equal cycles, respectively.} \label{fig:ah}
\end{figure}

\begin{figure}[htbp]
\includegraphics[scale=0.6,width=7.0cm]{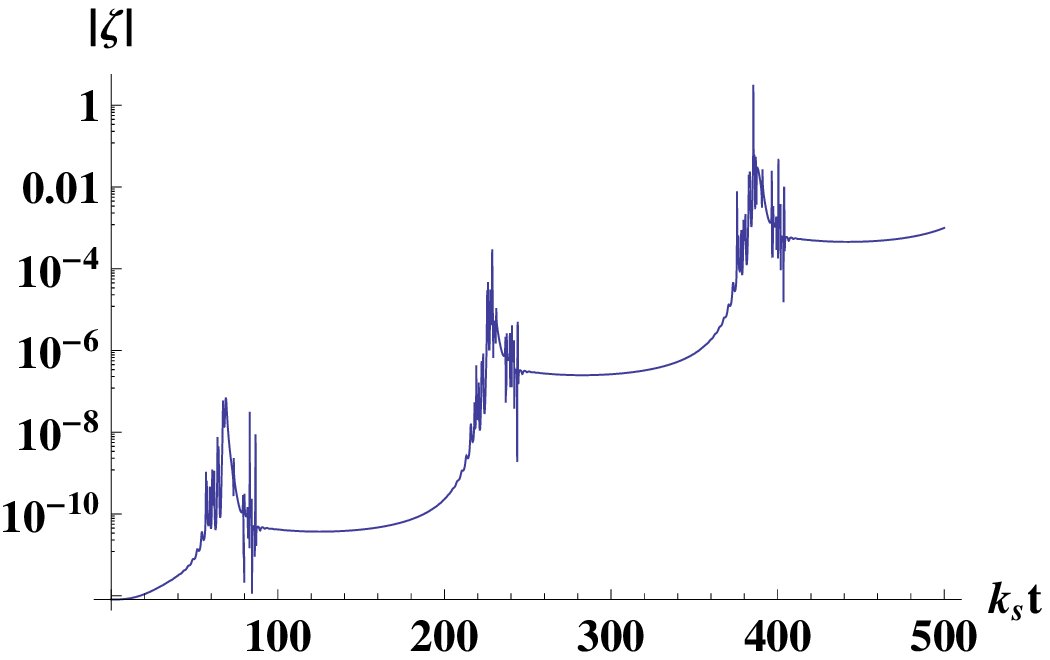}
\includegraphics[scale=0.6,width=7.0cm]{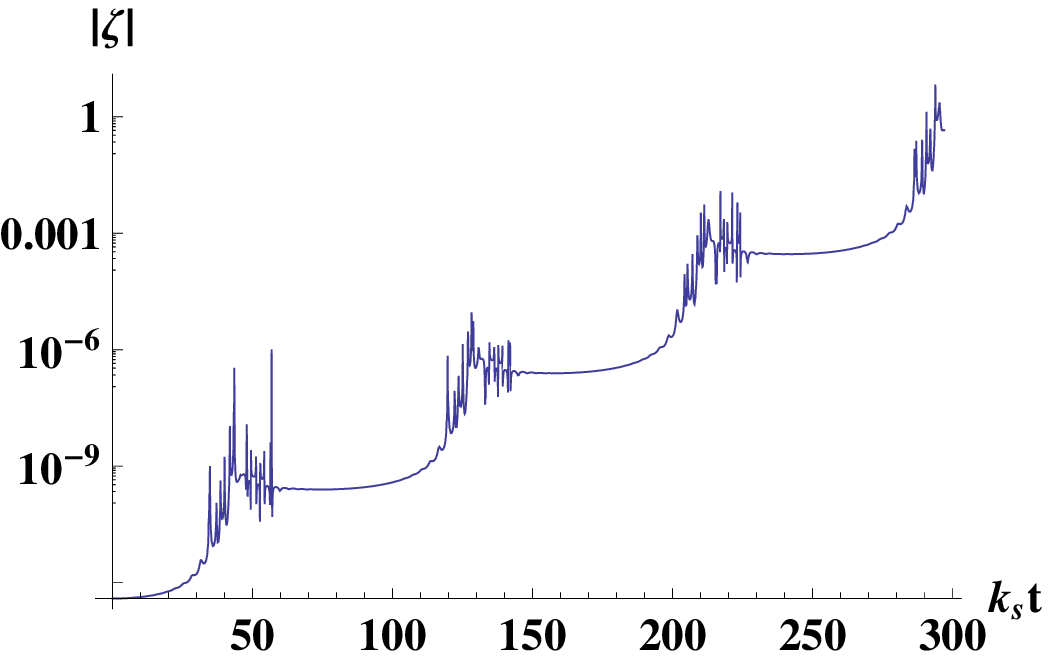}
\caption{ The upper panel is the evolution of  $\zeta_k$ in the
model with increasing cycles, and the lower panel is that in the
model with equal cycles. We can see that $\zeta_k$ is increased
cycle by cycle outside of Hubble scale, independent of the style of
cyclic models.} \label{fig:equl}
\end{figure}

The amplification for the perturbation modes, which are outside of
Hubble scale in the $j^{th}$ and $j^{th}+1$ cycles all along, can
be estimated as follows. In the $j^{th}$ cycle, after the bounce,
$\zeta_k$ is given by Eq.(\ref{zeta}), which will be unchanged up
to the end of the $j^{th}$ cycle. Then the universe enters into
the $j^{th}+1$ cycle, during the contraction of the $j^{th}+1$
cycle, $\zeta_k$ will continue to increase. Thus at certain time
during the contraction, in term of Eq.(\ref{zeta}), we have \ba
\zeta_k^{j+1}(t^{j+1})&\simeq &
\left(\frac{t_{B}^{j+1}-t^{j+1}}{t_{B}^{j+1}-t_{Ci}^{j+1}}\right)^{1-3n_{j+1}}\zeta_k^{j}(t_{Ce}^j)\nonumber\\
&\sim &
\left(\frac{t_{B}^{j+1}-t^{j+1}}{t_{B}^{j+1}-t_{Ci}^{j+1}}\right)^{1-3n_{j+1}}
\left(t_{B}^j-t_{Ce}^j\right)^{1-3n_{j}}
\nonumber\\
&\sim &  \exp{\left[{(3n^{j+1}-1){\cal N}^{j+1}\over
1-n^{j+1}}\right]}(H_{Ce}^j)^{3n_j-1}
, \label{zeta1}\ea since $H\sim 1/(t_B-t)$, which means that for
these modes still staying outside of Hubble scale the amplitude of
the spectrum will be amplified with same rate after each cycle. In
the third line, $ {\cal N}^{j+1}=\ln({a^{j+1}H^{j+1}\over
a_{Ci}^{j+1}H_{Ci}^{j+1}})$, which is \be {\cal N}^{j+1}\simeq
(1-n^{j+1})\ln\left({H^{j+1}\over H_{Ci}^{j+1}}\right), \ee is the
efolding number for the primordial perturbation generated during the
contraction of the $j^{th}+1$ cycle. Thus for the $j^{th}$ cycle,
the amplification of the perturbation amplitude is $\sim \exp{\cal
N}^j$. Thus with Eq.(\ref{zeta1}), we have

\ba \zeta_k^{j+1}(t^{j+1}) \simeq
\zeta_k^{1}(t_{Ce}^1)\left(\frac{t_{B}^{j+1}-t^{j+1}}{t_{B}^{j+1}-t_{Ci}^{j+1}}\right)^{1-3n_{j+1}}
\nonumber\\\prod_{s=1,...,j}\left(\frac{t_{B}^{s}-t_{Ce}^{s}}{t_{B}^{s}-t_{Ci}^{s}}\right)^{1-3n_{s}}
, \label{zeta2} \ea where $\zeta^1_k$ is the perturbation generated
during the first cycle. Eq.(\ref{zeta2}) indicates that the
curvature perturbation is multipled through cycles, and thus the
rate of the amplification is quite rapid, which is general. We have
assume that the linear perturbation approximation is satisfied all
along. However, it might be broken after one or some cycles, which
will bring a cutoff for $j$.

\begin{figure}[htbp]
\includegraphics[scale=0.6,width=6.0cm]{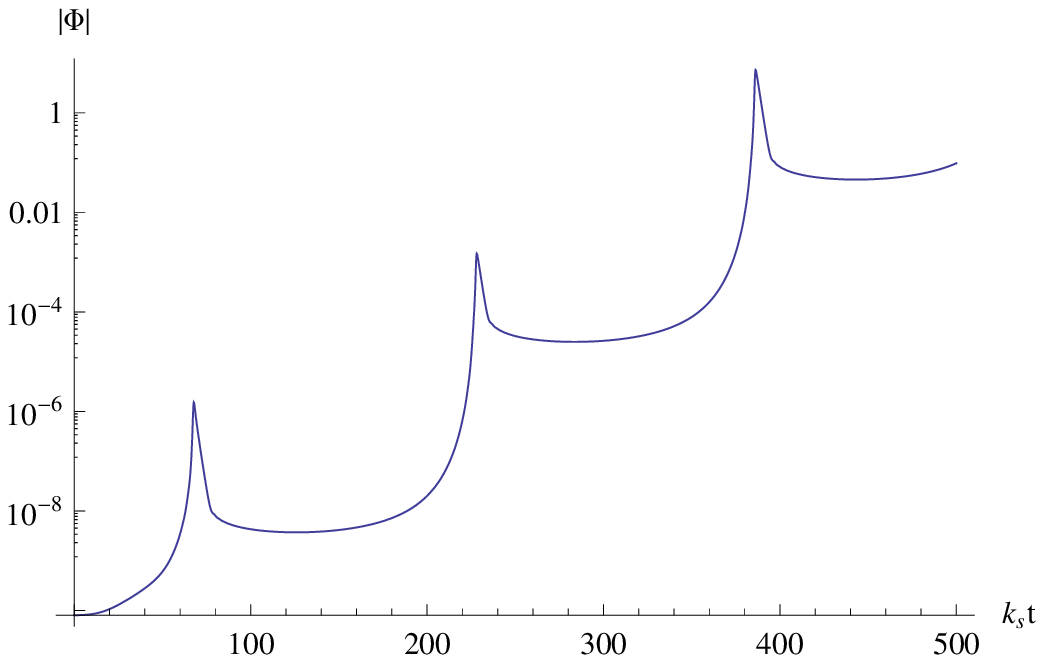}
\includegraphics[scale=0.6,width=6.0cm]{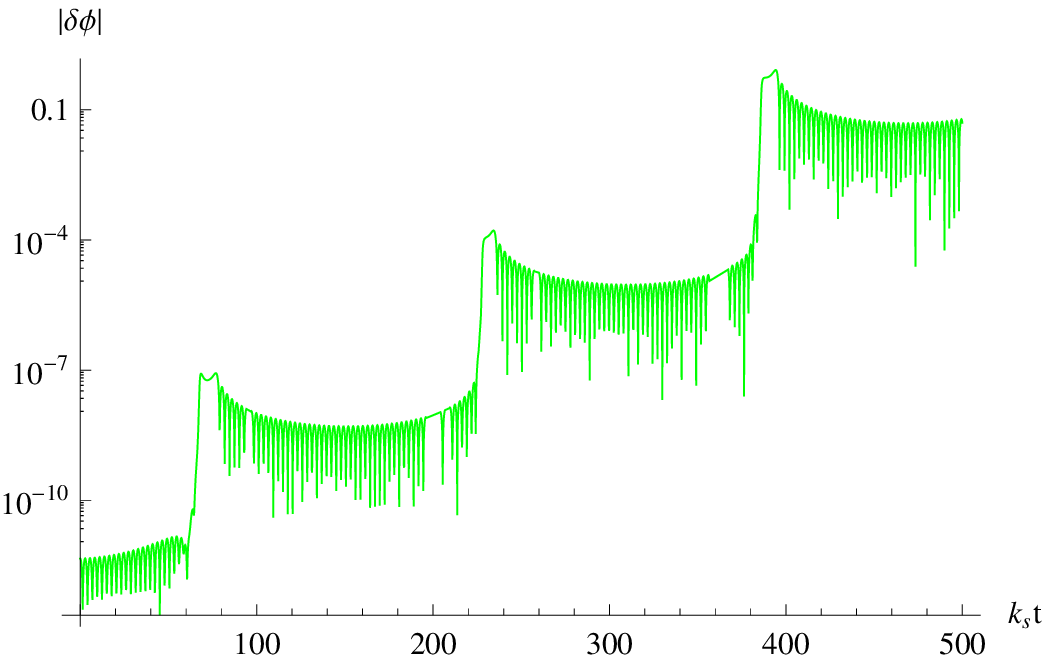}
\caption{ The evolution of $\Phi_k$ and $\delta\phi_k$ in the model
with increasing cycles. We can see that $\delta\phi_k$ is increased
in contracting phase and constant in expanding phase, thus the net
result is increased cycle by cycle. The corresponding mode is $k_s$
} \label{fig:incrphi}
\end{figure}

In Fig.4, the evolutions of the perturbation modes $\zeta_k$ is
plotted with the backgrounds in Fig.1 by numerically solving
Eqs.(\ref{u1}),(\ref{u2}),(\ref{u3}). We can clearly see that
$\zeta_k$ is increasing during the contraction and is constant
during the expansion, thus the net result is that the amplitude of
perturbation is amplified cycle by cycle, which is consistent with
Eq.(\ref{zeta2}). The metric perturbation $\Phi$ and the
perturbation $\delta\phi$ of background scalar field $\phi$ are
plotted in Figs.5 and 6. They obviously also increase cycle by
cycle. The results are consistent with above discussions, and also
the numerical results in \cite{AW},\cite{Cai0810},\cite{CK} for
one cycle, and also \cite{B0905}.

\begin{figure}[htbp]
\includegraphics[scale=0.6,width=6.0cm]{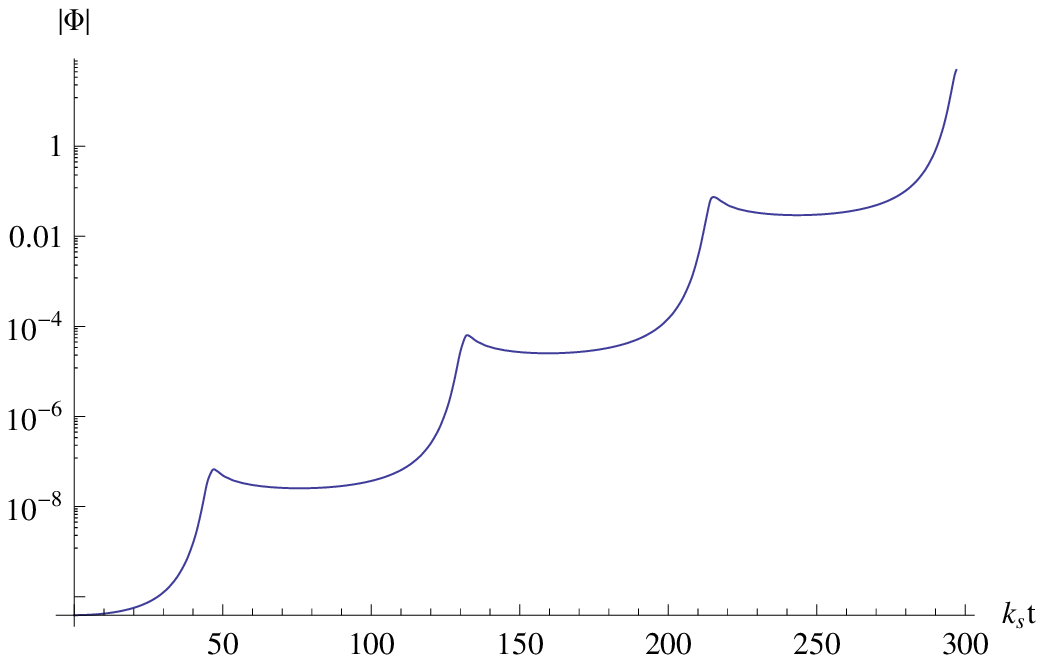}
\includegraphics[scale=0.6,width=6.0cm]{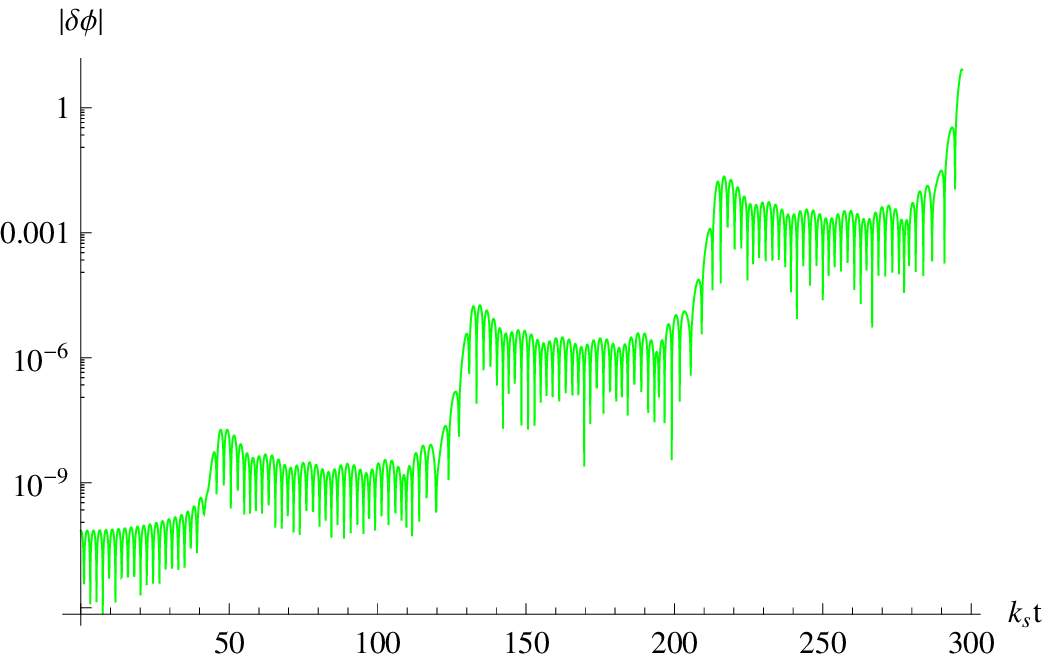}
\caption{ The evolution of $\Phi_k$ and $\delta\phi_k$ in the model
with equal cycles. Their characters are similar with increasing
cyclic models. The corresponding mode is $0.01k_s$ }
\label{fig:equlphi}
\end{figure}

\section{The evolution of isocurvature perturbations}

We will study the evolution of isocurvature perturbation. In
principle, there may be a set of scalar fields together
participating in the evolution of cyclic universe. The curvature
perturbation is the perturbation along the field trajectory in
fields space, while the isocurvature perturbation is the
perturbation orthogonal to the field trajectory \cite{GWBM}. The
general case is slightly involved. However, for simplicity, we
assume that there is only a field $\chi$ other than $\phi$, which
has nothing to do with the background evolution, and thus the
background field $\phi$ is responsible for the curvature
perturbation, while $\chi$ only contributes the isocurvature
perturbation.

The evolution of $\chi$ field is given by \be\label{chi}
\ddot\chi+3H\dot\chi+V_{\chi}= 0.\ee The simplest case is
$V(\chi)=0$, we have $\dot\chi\sim{1/a^3}$. Thus if initially $\dot
\chi\neq 0$, $\dot \chi$ will increase in the contracting phase,
while $\dot \chi\rightarrow 0$ rapidly in the expanding phase.

During the contraction phase (\ref{ac}) of each cycle, \be \chi \sim
\int {dt\over a^3} \sim(t_B-t)^{1-3n}. \ee Thus $\chi$ will farther
and farther deviate from its initial value in the model with equal
cycles, while will eventually stop in certain value in the models
with increasing cycle.

Alternatively, the potential is \be
V(\chi)=\frac{1}{2}M_\chi^2\chi^2. \label{m2}\ee

During the contraction, the solution of Eq.(\ref{chi}) is
approximately \be
\chi\simeq\frac{M_{p}}{\sqrt{3\pi}M_\chi(t_B-t)}\sin[M_\chi(t_B-t)].
\label{chim2}\ee The field $\chi$ will oscillate around its minimum
$\chi=0$ and its amplitude will increase during the contraction of
each cycle.

The perturbation equation of $\chi$ field is given by
Eq.(\ref{u1}), \be\label{chimass}
\delta\ddot\chi_k+3H\delta\dot\chi_k+(\frac{k^2}{a^2}+V_{\chi\chi})\delta\chi_k=
4\dot\Phi_k\dot\chi-2\Phi_k V_{\chi}. \ee

\subsection{The evolution of $\delta\chi$ without the coupling to the metric perturbation}

We firstly neglect the coupling of $\chi$ to the metric perturbation
$\Phi$, which is right only when the effect of $\chi$ on the
background evolution is completely negligible. When $V_{\chi\chi}\ll
H^2$, the equation of $\delta \chi$ is  \be
\delta\ddot\chi_k+3H\delta\dot\chi_k+\frac{k^2}{a^2}\delta\chi_k=0,
\label{dechi} \ee We define $\upsilon=a\delta\chi$, Eq.(\ref{dechi})
becomes \be\label{u}
\upsilon_k^{\prime\prime}+(k^2-\frac{a^{\prime\prime}}{a})\upsilon_k=0,
\ee where the prime is the derivative for the conformal time $\eta$.
Initially, the perturbation is deeply inside its Hubble scale,
$k^2\gg a^{\prime\prime}/{a}$, the solution is $ \upsilon_k\sim
\frac{1}{\sqrt{2k}} e^{-ik\eta}$. During the contraction, the
perturbation is extended outside of the Hubble scale. When $k^2\ll
a^{\prime\prime}/{a}$, the solution of Eq.(\ref{u}) can be written
as
 \be\label{chinom} \delta\chi_k\simeq C_{\delta\chi,2}\int\frac{d\eta}{a^2}+
C_{\delta\chi,1}\ee where the $C_{\delta\chi,1}$ mode is the
constant mode and the $C_{\delta\chi,2}$ mode is the mode dependent
on time. The spectrum of $\delta\chi_k$ will be determined by which
one of them is dominated. This solution is almostly same as
(\ref{c12}) of curvature perturbation. The difference is that for
$\zeta$, the $C_{\zeta,2}$ mode is $\sim \int d\eta/z^2$, which
equals $\int d\eta/a^2$ only when $\omega$ is constant. However,
around the bounce time, $\omega$ is rapidly changed.

The increasing of the amplitude of $\delta\chi_k$ during the
contraction is determined by the increasing mode \be
\delta\chi_k\sim \int {dt\over a^3} \sim {1\over (t_B-t)^{3n-1}}.
\label{chi2}\ee Thus $\delta\dot\chi_k\sim 1/a^3$. The matching
condition around the bounce is that both $\delta\chi_k$ and its time
derivative are continuous. Thus after the bounce
${\delta\dot\chi_k}$ will inherit the corresponding value before the
bounce. However, since during the expansion $\delta\dot\chi_k\sim
1/a^3$ is decreased. Thus before the contraction in following cycle,
its initial value is dependent on the ratio of the expansion of $a$
to the contraction in previous cycle. In general, we have \be {
\delta\dot\chi}^{j+1}_{Ee}=\left({a^{j}_{Ee}\over
a^{j+1}_{Ee}}\right)^3{\delta\dot \chi}^{j}_{Ee}\sim
\prod_{i=1,...,j} \left({a^{i}_{Ee}\over a^{i+1}_{Ee}}\right)^3{
\delta\dot\chi}^{1}_{Ee}. \label{dotdeltachi}\ee Thus if the cycle
is equal, the initial value of ${\delta\dot\chi_k}$ is same for each
cycle, however, if the cycle is increasing, it is decaying with the
increasing of the number of cycles.

The initial value of ${\delta\dot\chi_k^{j+1}}$ at the beginning
time of the contraction determines the amplitude of the increasing
of $\delta\chi_k^{j+1}$ in the $j^{th}+1$ cycle. Thus after the
bounce of the $j^{th}+1$ cycle, we generally have
 \be
\delta\chi_k^{j+1}\sim \left(1+\sum_{i=1,...,j}\int {\delta\dot
\chi_k^{i+1}} dt\right)\delta\chi_k^1, \label{chik2}\ee
where$\delta\chi_k^1$ is the perturbation generated during the first
cycle. Eq.(\ref{chik2}) indicates that for the model with equal
cycles, the perturbation modes will be amplified cycle by cycle and
the amplification rate at each cycle is same, however, for the model
with increasing cycles, since $\delta\dot\chi\sim 1/a^3$ is decaying
with the increasing of the number of cycles, the integration $\int
{\delta \dot\chi_k^{j+1}} dt$ will be negligible after one or some
cycles, which means that the perturbation $\delta\chi_k$ will be
amplified in the beginning and approach a constant with the
increasing of the number of cycles.

\begin{figure}[htbp]
\includegraphics[scale=0.6,width=7.0cm]{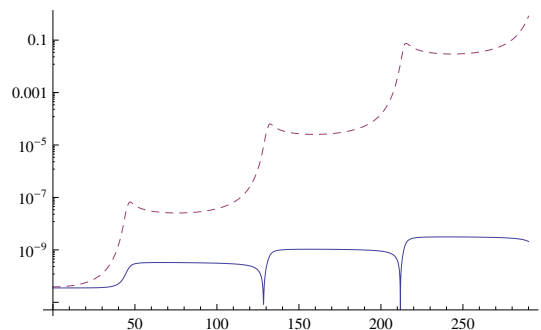}
\caption{ The solid line is the evolution of $\delta\chi_k$ in the
model with equal cycles, while the dashed line is the evolution of
$\Phi_k$.} \label{fig:eqm}
\end{figure}

\begin{figure}[htbp]
\includegraphics[scale=0.6,width=7.0cm]{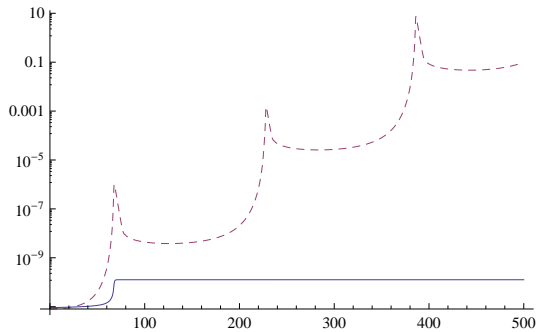}
\caption{ The solid line is the evolution of $\delta\chi_k$ in the
model with increasing cycles, while the dashed line is the
evolution of $\Phi_k$.} \label{fig:inm}
\end{figure}

The evolutions of $\delta\chi_k$ and $\Phi_k$ in the models with
equal cycles and increasing cycles are plotted in Figs.\ref{fig:eqm}
and \ref{fig:inm}, respectively. In Fig.\ref{fig:eqm},
$\delta\chi_k$ will increase cycle by cycle, and its increasing is
slower than that of $\Phi_k$, which can be actually noticed by
comparing Eqs.(\ref{zeta2}) and (\ref{chik2}). In Fig.\ref{fig:inm},
$\delta\chi_k$ will only increase at the first cycle and then tends
to constant in the following cycles. The numerical results are
consistent with the analytic estimates.

\subsection{The evolution of $\delta\chi$ with the coupling to the metric perturbation}

During the contraction of each cycle, the energy of $\chi$ field
${\dot\chi}^2\sim 1/a^6$ is rapidly increased. Its increasing is
generally faster than that of the background field $\phi$, thus it
might be required to include the coupling of $\chi$ to the metric
perturbation $\Phi$. The metric perturbation is amplified cycle by
cycle, which is faster than that of free $\delta\chi$, and thus
will source the evolution of $\delta\chi$. We will check it.

When $V_{\chi}=0$, that left in the right side of Eq.(\ref{chimass})
is $4\dot\Phi_k\dot\chi$, Eq.(\ref{chimass}) becomes
\be\label{chimetr}
\delta\ddot\chi_k+3H\delta\dot\chi_k+\frac{k^2}{a^2}\delta\chi_k=4\dot\Phi_k\dot\chi,
\ee In general, ${\dot \Phi}\simeq H \Phi$ and $\dot\chi\simeq
H\chi$. What we care is the evolution of perturbation on large
scale, i.e. $k^2\ll a^2H^2$. Thus Eq.(\ref{chimetr}) becomes
approximately \be H^2\delta\chi_k\sim H\Phi_k\dot\chi,\ee and thus
\be \delta\chi_k\sim\frac{\dot\chi}{H}\Phi_k. \label{a3H}\ee

In the model with equal cycles, $a$ is periodically equal. In
contracting phase of each cycle, we have \be
\delta\chi_k\sim\frac{1}{a^3H}\Phi_k\sim\frac{1}{(t_B-t)^{3n-1}}\Phi_k
\ee where we have used $a\sim(t_B-t)^n$ and $H\sim1/(t_B-t)$. We
can conclude that during the contraction with $n>\frac{1}{3}$, the
increasing of the amplitude of $\delta\chi_k$ is faster than that
of the metric perturbation $\Phi_k$.

In the model with increasing cycles, $\dot\chi$ will become quite
small after some cycles, since the maximal value of $a$ in each
cycle is increased with the increasing of the number of cycles.
Thus after one or some cycles, $\Phi_k$ will hardly have effect on
$\delta\chi_k$, therefore $\delta\chi_k$ will approach a constant.

Alternatively, for the potential (\ref{m2}), the equation of
$\delta\chi$ is \be
\delta\ddot\chi_k+3H\delta\dot\chi_k+(\frac{k^2}{a^2}+M_\chi^2)\delta\chi_k=4\dot\Phi_k\dot\chi-2\Phi_kM_\chi^2\chi
\ee When $M^2_\chi\ll H^2$, the result will be same with that of
$V=0$. While for $M^2_\chi\gg H^2$, we have \be\label{solchi}
 \delta\chi_k\sim\chi \Phi_k, \ee where $\chi$ is given by
 Eq.(\ref{chim2}). Thus
the solution (\ref{solchi}) is approximately
\be
\delta\chi_k\sim\frac{M_{p}}{M_\chi(t_B-t)}\sin[M_\chi(t_B-t)]\Phi_k.
\ee
Thus the perturbation will oscillate and its amplitude will be
amplified faster than $\Phi_k$. However, the increasing rate of
the amplitude dose not decay with the increasing of the number of
cycles.

\begin{figure}[htbp]
\includegraphics[scale=0.6,width=7.0cm]{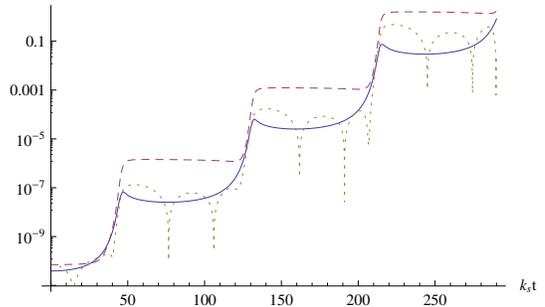}
\caption{ The background is that of the model with equal cycles.
The solid line is the evolution of $\Phi_k$ and the dashed line
and the dotted line are the evolution of $\delta\chi_k$ with
$V(\chi)=0$ and $V(\chi)={1\over 2}M_{\chi}^2\chi^2$,
respectively. } \label{fig:eqcm}
\end{figure}

\begin{figure}[htbp]
\includegraphics[scale=0.6,width=7.0cm]{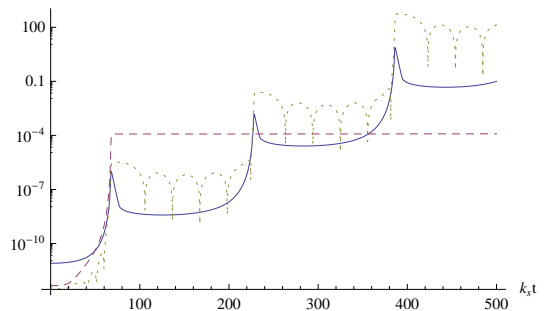}
\caption{The background is that of the model with increasing
cycles. The solid line is the evolution of $\Phi_k$ and the dashed
line and the dotted line are the evolution of $\delta\chi_k$ with
$V(\chi)=0$ and $V(\chi)={1\over 2}M_{\chi}^2\chi^2$,
respectively. } \label{fig:incm}
\end{figure}

The numerical results for the models with equal cycles and
increasing cycle, respectively, are plotted in Figs.\ref{fig:eqcm}
and \ref{fig:incm}. We can see that the perturbation
$\delta\chi_k$ is amplified faster than $\Phi_k$ cycle by cycle.
The oscillating of $\chi$ leads the oscillating of $\delta\chi_k$.
These results are consistent with analytical discussions.

\section{Discussion }

The cosmological cyclic scenario offered an alternative insight
into the origin of observable universe. In general, it is thought
that the background evolution of cyclic universe is globally
homogeneous cycle by cycle. However, the amplitude of curvature
perturbation on large scale will be amplified cycle by cycle.

In certain sense, the amplification of curvature perturbation
might eventually destroy the homogeneity of background, which
might lead to the ultimate end of cycles of global universe.
However, it can be argued \cite{Piao0901,Piao1001} that for the
model with increasing cycles, the global universe will possibly
evolve into a fissiparous multiverses after one or some cycles, in
which the cycles will continue only at corresponding local
regions, inside of which the background is homogeneous.

The amplitude of the isocurvature perturbation on large scale may
be also amplified cycle by cycle. However, the rate of its
amplification is slower than that of curvature perturbation, and
further for the model with increasing cycles, with the increasing
of the number of cycles the amplification of its amplitude will be
suppressed, and after one or some cycles the resulting amplitude
will approach a constant. However, if the coupling to the metric
perturbation is included, the result will be altered. This might
have interesting implication to the application of the curvaton
mechanism
\cite{Lyth:2001nq},\cite{Enqvist:2001zp},\cite{Moroi:2001ct},\cite{Linde:1996gt},
called the bounce curvaton scenario \cite{Cai:2011zx}.

\textbf{Acknowledgments} We thank Jun Zhang for discussion, and
Yi-Fu Cai, Taotao Qiu for comments of earlier draft. This work is
supported in part by NSFC under Grant No:10775180, 11075205, in
part by the Scientific Research Fund of GUCAS(NO:055101BM03), in
part by National Basic Research Program of China, No:2010CB832804.

\end{document}